\ifx\pdfoutput\undefined
\documentstyle[prl,aps,amsmath,amssymb,eucal,multicol]{revtex}
\else
\documentstyle[prl,aps,amsmath,amssymb,eucal,multicol,hyperref]{revtex}
\hypersetup{%
  pdfpagemode=None,
  pdftitle=UAAHS,
  pdfauthor=J.~Benoit and R.~Dandoloff,
  pdfsubject=math-ph/0004029,
  pdfstartpage=1,
  pdfview=FitBH,
  pdfstartview=FitBH,
  citecolor=green,
  urlcolor=blue,
  pdfhighlight=/N,
  backref=false,
  pagebackref=false,
  }
\fi

\newcommand{\ci}{\qi}
\newcommand{\cw}{{\mathsf w}}
\newcommand{\qi}{{\mathsf i}}
\newcommand{\gtse}{{\mathrm e}}
\newcommand{\Kronecker}{\delta}
\newcommand{\Planck}{\hslash}
\newcommand{\gtsJ}{{\mathit J}}
\newcommand{\gtsHm}{{\EuScript H}}
\newcommand{\gtsDimension}{{\mathrm d}}
\newcommand{\gtsastr}{{\mathrm e}}
\newcommand{\gtsVertexI}{i}
\newcommand{\gtsVertexJ}{j}
\newcommand{\gtsSpin}{{\mathit S}}
\newcommand{\gtsSpinDim}{{\mathit s}}
\newcommand{\gtsLStep}{{\mathsf a}}
\newcommand{\gtsLSpaceGroup}{{\mathsf g}}
\newcommand{\gtsLColourPermutation}{\gamma}
\newcommand{\gtsSpinNorm}{\nu}
\newcommand{\gtsAFOP}{{\mathit n}}
\newcommand{\gtsAFTR}{{\mathit l}}
\newcommand{\gtsAFAlg}{{\mathit d}}
\newcommand{\gtsCubicPolytope}{{\mathsf C}}
\newcommand{\gtsTriPolytope}{{\mathsf T}}
\newcommand{\gtsHVertexI}{p}
\newcommand{\gtsGP}{\gamma}
\newcommand{\gtsAFChirality}{\chi}
\newcommand{\gtsAFNChirality}{\widehat{\chi}}

\begin{document}

\draft
\title{A Uniform Approach to Antiferromagnetic Heisenberg Spins\\
    on Low Dimensional Lattices%
  \thanks{\textsc{Physics Letters A} \textbf{276}, 175 (2000) %
  [\texttt{math-ph}/0004029]}%
  }
\author{%
  J.~Benoit%
  \thanks{E-mail: jerome.benoit@ptm.u-cergy.fr}%
  and
  R.~Dandoloff%
  \thanks{E-mail: rossen.dandoloff@ptm.u-cergy.fr}%
  }
\address{
    Laboratoire de Physique Th\'eorique et Mod\'elisation (\textsc{CNRS-ESA}~8089 ),\\
    Universit\'e de Cergy-Pontoise,
    5~Mail~Gay-Lussac,
    95031~Cergy-Pontoise, France}
\date{Received 26 June 2000; %
  accepted 22 September 2000\\
  Communicated by A.~R. Bishop}
\maketitle
\begin{abstract}
Using group theoretical methods we
show for both the triangular and square lattices
that in the continuum limit
the antiferromagnetic order parameter lives on $\mathrm{SO}(3)$
without respect of the initial lattice.
For the antiferromagnetic chain we recover the Haldane decomposition.
This order parameter interacts with a local gauge field
rather than with a global one
as implicitly suggested in the literature
which in our approach appears in a rather natural manner.
In fact this merely corresponds to
a novel extension of the spin group by a local gauge field.
This analysis based on the real division algebras
$\mathbb{A}_{\gtsDimension}$
applies to low dimensional lattices
($\gtsDimension=1,2,3,4$).
\newline\copyright 2000 Elsevier Science B.~V. All~rights~reserved.
\end{abstract}


\begin{multicols}{2}
Since the discovery of the high temperature superconductivity
bidimensional antiferromagnetic Heisenberg models and their continuum
limit have attracted considerable interest
\cite{DombreReadA,DombreReadT,TTQHA,SGSBA,ShankarRead}.
The problem of the continuum limit is directly related to
the presence or not of a Hopf term
in the Lagrangian of the bidimensional
antiferromagnetic Heisenberg model
\cite{DPWXXX,FBT,PPFBT}.
This problem is still open
\cite{BBDXXX,AMDQHA}.

The continuum limit of
the bidimensional antiferromagnetic Heisenberg model
is much more difficult to achieve than may be expected
in particular for the triangular and hexagonal lattices.
This difficulty has led the authors to outline the construction
of novel local antiferromagnetic degrees of freedom
for low dimensional antiferromagnetic lattices.
This new construction naturally reproduces
the real division algebra hierarchy \cite{Ebbinghaus,Dixon,duVal}
satisfied by the nonlinear $\sigma$-model \cite{PPFBT,GurseyTze,RealAlgebra}
(which is accepted to correspond to the continuum limit
of the Heisenberg Hamiltonian for classical
isotropic antiferromagnets in two dimensions
\cite{BelavinPolyakov,Trimper,Chakravarty,HaldanePRL,Fradkin}).
Real division algebras are known as powerful tools to describe
regular tessellations in low dimension
(the complex algebra $\mathbb{C}$ on plane
and the quaternion algebra $\mathbb{H}$ in space \cite{duVal,RCPCoxeter})
and interactions in particle physics \cite{Dixon,GurseyTze}.
This powerfulness is employed
to introduce effortlessly new degrees of freedom.
This algebraic construction suggests
an original microscopic mechanism
and reveals the presence of a novel local gauge field
which is expected to give rise
to a Hopf term in the continuum limit\cite{HSTL}.
In the present paper we will focus
on the antiferromagnetic chain,
the antiferromagnetic triangular and square lattices
only.

First,
let us consider compact regular tessellations
partitioning low dimensional Euclidian space
($\gtsDimension=1,2,3,4$)
with identical regular polytopes\cite{Coxeter}.
Then each crystal will be characterized
by a regular polytope:
the crystal will have the dimension $\gtsDimension$
of the regular polytope,
the symmetries which leave the regular polytope invariant
will leave the crystal invariant,
the translations which reproduce the crystal
with a single regular polytope
will leave the crystal coinciding with itself,
and
the length of the regular polytope edge
will correspond to the crystal step $\gtsLStep$.
Implicitly we have associated
to each crystal site $\left(\gtsVertexI\right)$
a spin operator ${\gtsSpin_{\!\gtsVertexI}}^{\alpha}$
which generate a Lie algebra\cite{Sattinger}
according to the commutation relation
\begin{equation}
\label{Spin/commutation/crude}
\left[{\gtsSpin_{\!\gtsVertexI}}^{\alpha},
{\gtsSpin_{\!\gtsVertexJ}}^{\beta}\right]=
\ci\Planck\:
\gtsastr^{\alpha\beta\gamma}\Kronecker_{\gtsVertexI\gtsVertexJ}\;
{\gtsSpin_{\!\gtsVertexI}}_{\gamma}
\end{equation}
with usual notations.
Therefore the Casimir operator $\gtsSpin^{2}$
fixes the dimension $\gtsSpinDim$ of the local degree of freedom:
\begin{equation}
\label{Spin/Casimir/crude}
\gtsSpin^{2}\equiv%
  {\gtsSpin_{\!\gtsVertexI}}^{\alpha}%
  {\gtsSpin_{\!\gtsVertexI}}_{\alpha}=%
    \gtsSpinDim\left(\gtsSpinDim+1\right)
\qquad\forall\gtsVertexI.
\end{equation}
Long distance magnetic behaviour may be tackled
applying a prescription due to Affleck \cite{SGSBA,QSCHG}:
one tends conjointly the step $\gtsLStep$
and the quantum number $\gtsSpinDim$
to zero and infinity respectively
while maintaining constant measurable physical entities.
In this respect spins are implicitly treated
in the classical limit
since the commutation relation (\ref{Spin/commutation/crude}),
which to have a physical sense must be written as
\begin{equation}
\label{Spin/commutation/nested}
\left[\frac{{\gtsSpin_{\!\gtsVertexI}}^{\alpha}}{\gtsSpinDim},
\frac{{\gtsSpin_{\!\gtsVertexJ}}^{\beta}}{\gtsSpinDim}\right]=
\ci\frac{\Planck}{\gtsSpinDim}\:
\gtsastr^{\alpha\beta\gamma}\Kronecker_{\gtsVertexI\gtsVertexJ}\;
\frac{{\gtsSpin_{\!\gtsVertexI}}_{\gamma}}{\gtsSpinDim},
\end{equation}
vanishes when $\gtsSpinDim$ approaches infinity.
Let us now take as
effective Hamiltonian describing exchange interactions
the following antiferromagnetic Heisenberg Hamiltonian
\begin{equation}
\label{Hm/Heiseinberg/afnn}
\gtsHm_{H}=
\gtsJ\sum\limits_{\left<\gtsVertexI,\gtsVertexJ\right>}
  {\gtsSpin_{\!\gtsVertexI}}^{\alpha}
  {\gtsSpin_{\!\gtsVertexJ}}_{\alpha},
\end{equation}
where $\gtsJ$ is the positive coupling constant
between nearest-neighbour spins
$\left<\gtsVertexI,\gtsVertexJ\right>$.

At high temperatures (above the Curie temperature)
each spin ${\gtsSpin_{\!\gtsVertexI}}^{\alpha}$
is oriented at random:
the symmetry of the spin crystal
is described by the crystal space group
$\left(\gtsLSpaceGroup\right)$ ---
combinations of translations and rotations
that leave the crystal unchanged.
This corresponds to the paramagnetic phase.
At very low temperatures (near the absolute zero)
neighbouring spins will behave according to
exchange interactions:
a spin arrangement may arise in which
the spins form a repeating pattern.
Rather,
a local antiferromagnetic symmetry emerges
as a pattern of oriented spins.
Symbolizing each local orientation by colour,
the symmetry of the spin crystal may be
described by the crystal colour group
$\left(\gtsLSpaceGroup,\gtsLColourPermutation\right)$ ---
combinations of translations and rotations
and permutations of the colours
that leave the coloured crystal invariant
\cite{SPNG,CSSchwarzenberger,CGMS,TCSPC}.
This antiferromagnetic symmetry characterizes
antiferromagnetic phases.

From now on,
let us restrict ourselves to the permutation group
$\left(\gtsLColourPermutation\right)$
to which the group of transformations
preserving the regular polytope is homomorphic.
Thinking of the regular polytope
as a regular tessellation of the sphere $S^{\gtsDimension-1}$
suggests to regard the coloured crystal
as a crystal of spheres $S^{\gtsDimension-1}$,
each consistently adorned with a coloured regular polytope;
we will call hyper-crystal such a crystal
and hyper-site each adorned sphere $S^{\gtsDimension-1}$.
By construction the hyper-crystal is colour-blind,
thereby the hyper-crystal appears as
the crystal of the antiferromagnetic phase:
a relevant local order parameter describing
the antiferromagnetic phase
has to be associated to a hyper-site
and a relevant continuum limit passage procedure
has to shrink the hyper-crystal
\cite{DombreReadA,DombreReadT,RGSHM}.
A basic approach consists of computing
the new local degree of freedom associated to
a hyper-site as being linear combinations
of the old local degrees of freedom (namely the spins)
that reduce the irreducible representations of
the transformation group
leaving the regular polytope invariant\cite{RGSHM}.

Noting that the transformation group leaving
a regular tessellation of the sphere $S^{\gtsDimension-1}$
invariant admits representations
in the real division algebra $\mathbb{A}_{\gtsDimension}$
of dimension~$2^{\gtsDimension-1}$
\cite{Ebbinghaus,Dixon,duVal,GurseyTze,RCPCoxeter},
we shall compute the linear combinations
in the algebra $\mathbb{A}_{\gtsDimension}$.
\textit{A priori},
we choose as local order parameter
the linear combination invariant under
the local antiferromagnetic symmetry
and normalized to unity in the classical limit.
Furthermore in the continuum limit we assume that
the remaining linear combinations behave
as if the spins of the regular polytope
were effectively described by a continuum field.
Ultimately we impose a local representation:
the new local degrees of freedom associated to
a hyper-site are defined up to a local rotation.
Because this local rotation does alter
the new degrees of freedom of the hyper-site,
this local rotation may be interpreted as
a local gauge field\cite{Naber,BailinLove}.
The latter hypothesis is the most novel:
the representation implicitly chosen in the literature
is global\cite{RGSHM}.

From this perspective,
exchange interactions do not generate
a global antiferromagnetic arrangement ``\`a la N\'eel''
but enforce spins to organize themselves in hyper-sites:
the spins belonging to a same hyper-site bind together
while dissociating from spins belonging to
nearest-neighbour hyper-sites.
In brief, each hyper-site acts as a hyper-particle.
Thus long distance behaviour is caused
by hyper-particle interactions described
as a dissociation between spins belonging to
nearest-neighbour hyper-particles.
Note that each hyper-particle corresponds to
a regular antiferromagnetic crystal
on the sphere $S^{\gtsDimension-1}$.
Furthermore the group theory allows us to separate
the antiferromagnetic symmetry due to exchange interactions
and the symmetry inherited from the regular polytope,
the crystal maintaining a certain cohesion
conveyed by the local gauge field.
Consequently the new local degrees of freedom
inherit the algebraic hierarchy
satisfied by the transformation groups
preserving the regular polytopes,
that is to say the real division algebra hierarchy
\cite{Ebbinghaus,Dixon,duVal}.
This inheritance does not appear so obvious
when we treat any regular antiferromagnetic system
no more as a hyper-crystal but as a coloured crystal:
this insight leads us to specify the colour attribute
as a fixed orientation (``\`a la N\'eel'')
defined up to a local gauge field.
In other words, we have extended
the spin group \cite{SPNG} by a local gauge field
and outlined an approach.
As far as we know,
such extension of the spin group
has not been carried out in the literature.
Below we show why these ideas are appealing.

To illustrate our approach, we first consider
the regular antiferromagnetic chain.
The regular chain corresponds to the line
filled with the $1$-cube $\gtsCubicPolytope^{0}$
represented in the real division algebra $\mathbb{R}$
by the pair
\begin{equation}
\label{rp/chain/alg}
\gtsCubicPolytope^{0}=\left\{+1,-1\right\}=S^{0}.
\end{equation}
The transformations preserving the $1$-cube $\gtsCubicPolytope^{0}$
are the identity $1$ and the rotation/antisymmetry $-1$.
For each hyper-site $\left(\gtsHVertexI\right)$
of an arbitrarily extracted hyper-chain,
let us denote by $\left(\gtsHVertexI_{0}\right)$
and $\left(\gtsHVertexI_{1}\right)$ the two sites
represented respectively by
the real numbers $1$ and $-1$.
\begin{subequations}
\label{afl/chain/AFDF}
Conforming to our prescription we compute
the local antiferromagnetic order parameter
${\gtsAFOP_{\gtsHVertexI}}^{\alpha}$ such that
\begin{align}
\label{afl/chain/AFDF/AFOP}
2\,\gtsSpinNorm{\gtsAFOP_{\gtsHVertexI}}^{\alpha}&=%
  {\gtsSpin_{\!\gtsHVertexI_{0}}}^{\alpha}
  -{\gtsSpin_{\!\gtsHVertexI_{1}}}^{\alpha},
\intertext{and the trivial representation
${\gtsAFTR_{\gtsHVertexI}}^{\alpha}$ such that}
\label{afl/chain/AFDF/AFTR}
2\,\gtsLStep{\gtsAFTR_{\gtsHVertexI}}^{\alpha}&=%
  {\gtsSpin_{\!\gtsHVertexI_{0}}}^{\alpha}
  +{\gtsSpin_{\!\gtsHVertexI_{1}}}^{\alpha}.
\end{align}
\end{subequations}
\begin{subequations}
\label{afl/chain/rel}
In the classical limit
the Casimir operator $\gtsSpin^{2}$ (\ref{Spin/Casimir/crude})
imposes between
the new local antiferromagnetic degrees of freedom
the following true relations:
\begin{gather}
\label{afl/chain/rel/unit}
\gtsAFOP_{\gtsHVertexI}^2=%
  1-{\scriptstyle\frac{\gtsLStep^2}{\gtsSpinNorm^2}}\;
  \gtsAFTR_{\gtsHVertexI}^2,\\
\label{afl/chain/rel/chirality}
{\gtsAFOP_{\gtsHVertexI}}^{\alpha}{\gtsAFTR_{\gtsHVertexI}}_{\alpha}=0,
\end{gather}
\end{subequations}
with $\gtsSpinNorm\equiv\sqrt{\gtsSpinDim\left(\gtsSpinDim+1\right)}$.
Clearly the constraints (\ref{afl/chain/rel}) show that
the number of degrees of freedom (equals~to~$4$) is conserved.
Moreover,
when the step $\gtsLStep$ and the quantum number $\gtsSpinDim$
tend conjointly to zero and infinity respectively
(Affleck prescription \cite{SGSBA,QSCHG}),
the order parameter ${\gtsAFOP_{\gtsHVertexI}}^{\alpha}$
is effectively normalized to unity.
Finally let us observe that in the continuum limit
the order parameter lives on the unit sphere $S^2$.
By reversing the relationships (\ref{afl/chain/AFDF}),
the internal spins of the hyper-particle $\left(\gtsHVertexI\right)$
read
\begin{subequations}
\label{afl/chain/AFDF/inv}
\begin{align}
\label{afl/chain/AFDF/inv/zero}
{\gtsSpin_{\!\gtsHVertexI_{0}}}^{\alpha}&=%
  \hphantom{-}\gtsSpinNorm\,{\gtsAFOP_{\gtsHVertexI}}^{\alpha}%
  +\gtsLStep\,{\gtsAFTR_{\gtsHVertexI}}^{\alpha},\\
\label{afl/chain/AFDF/inv/one}
{\gtsSpin_{\!\gtsHVertexI_{1}}}^{\alpha}&=
  -\gtsSpinNorm\,{\gtsAFOP_{\gtsHVertexI}}^{\alpha}%
  +\gtsLStep\,{\gtsAFTR_{\gtsHVertexI}}^{\alpha}.
\end{align}
\end{subequations}
Henceforth, as expected,
the Haldane decomposition \cite{HaldanePRL} is reproduced.

Next,
let us focus on the antiferromagnetic triangular lattice.
The triangular lattice corresponds to
the plane filled with the equilateral triangle $\gtsTriPolytope^{1}$
represented in the real division algebra $\mathbb{C}$
by the triplet
\begin{equation}
\label{afl/tri/rp/alg}
\gtsTriPolytope^{1}=\left\{1,\cw,\cw^2\right\},
\end{equation}
where $\cw\equiv\gtse^{\ci\,\frac{2\pi}{3}}$.
Combinations of complex conjugation and rotation $\cw$
bring the equilateral triangle $\gtsTriPolytope^{1}$
into coincidence with itself.
For any arbitrarily extracted hyper-crystal,
the three sites of any hyper-site $\left(\gtsHVertexI\right)$
represented respectively by
the complex numbers $1$, $\cw$ and $\cw^2$
are denoted respectively by
$\left(\gtsHVertexI_{0}\right)$,
$\left(\gtsHVertexI_{1}\right)$ and
$\left(\gtsHVertexI_{2}\right)$.
\begin{subequations}
\label{afl/tri/AFDF}
Since the antiferromagnetic symmetry
is fulfilled by the rotation~$\cw$,
the local antiferromagnetic order parameter
${\gtsAFOP_{\gtsHVertexI}}^{\alpha}$
satisfies
\begin{align}
\label{afl/tri/AFDF/AFOP}
{\scriptstyle\frac{3}{\sqrt{2}}}\,%
\gtsSpinNorm{\gtsAFOP_{\gtsHVertexI}}^{\alpha}&=%
  \gtse^{-\ci\,\gtsGP_{\gtsHVertexI}}%
  \left[%
    {\gtsSpin_{\!\gtsHVertexI_{0}}}^{\alpha}
    +\cw\,{\gtsSpin_{\!\gtsHVertexI_{1}}}^{\alpha}
    +\cw^2\,{\gtsSpin_{\!\gtsHVertexI_{2}}}^{\alpha}
  \right],
\intertext{whereas the trivial representation
${\gtsAFTR_{\gtsHVertexI}}^{\alpha}$ reads}
\label{afl/tri/AFDF/AFTR}
3\,\gtsLStep^2{\gtsAFTR_{\gtsHVertexI}}^{\alpha}&=%
  {\gtsSpin_{\!\gtsHVertexI_{0}}}^{\alpha}
  +{\gtsSpin_{\!\gtsHVertexI_{1}}}^{\alpha}
  +{\gtsSpin_{\!\gtsHVertexI_{2}}}^{\alpha}.
\end{align}
\end{subequations}
Let us stress that
the complex representation (\ref{afl/tri/AFDF/AFOP})
is defined up to a local gauge field $\gtsGP_{\gtsHVertexI}$.
\begin{subequations}
\label{afl/tri/rel}
At the classical limit
the new local degrees of freedom verify
the following constraints:
\begin{gather}
\label{afl/tri/rel/unit}
{\gtsAFOP_{\gtsHVertexI}}^{\alpha}{^{\dag}\gtsAFOP_{\gtsHVertexI}}_{\alpha}=%
  1-{\scriptstyle\frac{\gtsLStep^4}{\gtsSpinNorm^2}}\;
  \gtsAFTR_{\gtsHVertexI}^2,\\
\label{afl/tri/rel/chirality}
{\gtsAFOP_{\gtsHVertexI}}^{\alpha}{\gtsAFOP_{\gtsHVertexI}}_{\alpha}=%
  -{\scriptstyle 2\sqrt{2}\:\frac{\gtsLStep^2}{\gtsSpinNorm}}\,%
  \gtse^{-\ci\,\gtsGP_{\gtsHVertexI}}\;%
  {\gtsAFTR_{\gtsHVertexI}}^{\alpha}{^{\dag}\gtsAFOP_{\gtsHVertexI}}_{\alpha}.
\end{gather}
\end{subequations}
Therefrom
the number of degrees of freedom (equals~to~$6$) is preserved
and the order parameter ${\gtsAFOP_{\gtsHVertexI}}^{\alpha}$
is effectively normalized to unity in the continuum limit.
Then it turns out to be advantageous to introduce
the local linear combinations
\begin{subequations}
\label{afl/tri/XY}
\begin{align}
\label{afl/tri/XY/X/def}
{{\gtsAFOP_{X}}_{\gtsHVertexI}}^{\alpha}&\equiv%
  {\scriptstyle\frac{2}{3\gtsSpinNorm}}\,
  \left[%
  {\gtsSpin_{\!\gtsHVertexI_{0}}}^{\alpha}%
  -{\scriptstyle\frac{1}{2}}\left({\gtsSpin_{\!\gtsHVertexI_{1}}}^{\alpha}%
    +{\gtsSpin_{\!\gtsHVertexI_{2}}}^{\alpha}\right)%
  \right],\\
\label{afl/tri/XY/Y/def}
{{\gtsAFOP_{Y}}_{\gtsHVertexI}}^{\alpha}&\equiv%
  {\scriptstyle\frac{2}{3\gtsSpinNorm}}\,
  {\scriptstyle\frac{\sqrt{3}}{2}}%
  \left({\gtsSpin_{\!\gtsHVertexI_{1}}}^{\alpha}%
    -{\gtsSpin_{\!\gtsHVertexI_{2}}}^{\alpha}\right);
\end{align}
\end{subequations}
in the classical limit we read
\begin{subequations}
\label{afl/tri/XY/rel}
\begin{gather}
\label{afl/tri/XY/rel/X}
{{\gtsAFOP_{X}}_{\gtsHVertexI}}^2=1%
  -{\scriptstyle 2\frac{\gtsLStep^2}{\gtsSpinNorm}}\,%
    {{\gtsAFOP_{X}}_{\gtsHVertexI}}^{\alpha}%
    {\gtsAFTR_{\gtsHVertexI}}_{\alpha}%
  -{\scriptstyle \frac{\gtsLStep^4}{\gtsSpinNorm^2}}\,%
    {\gtsAFTR_{\gtsHVertexI}}^{2},\\
\label{afl/tri/XY/rel/Y}
{{\gtsAFOP_{Y}}_{\gtsHVertexI}}^2=1%
  +{\scriptstyle 2\frac{\gtsLStep^2}{\gtsSpinNorm}}\,%
    {{\gtsAFOP_{X}}_{\gtsHVertexI}}^{\alpha}%
    {\gtsAFTR_{\gtsHVertexI}}_{\alpha}%
  -{\scriptstyle \frac{\gtsLStep^4}{\gtsSpinNorm^2}}\,%
    {\gtsAFTR_{\gtsHVertexI}}^{2},\\
\label{afl/tri/XY/rel/XY}
{{\gtsAFOP_{X}}_{\gtsHVertexI}}^{\alpha}%
  {{\gtsAFOP_{Y}}_{\gtsHVertexI}}_{\alpha}=
  {\scriptstyle 2\frac{\gtsLStep^2}{\gtsSpinNorm}}\,%
    {{\gtsAFOP_{Y}}_{\gtsHVertexI}}^{\alpha}%
    {\gtsAFTR_{\gtsHVertexI}}_{\alpha}.%
\end{gather}
\end{subequations}
As in the continuum limit the pair
$\left({{\gtsAFOP_{X}}_{\gtsHVertexI}}^{\alpha},%
  {{\gtsAFOP_{Y}}_{\gtsHVertexI}}^{\alpha}\right)$
becomes an orthonormal dihedron,
let us bring in the new operator
${{\gtsAFOP_{Z}}_{\gtsHVertexI}}^{\alpha}$
such that the triplet
$\left({{\gtsAFOP_{X}}_{\gtsHVertexI}}^{\alpha},%
  {{\gtsAFOP_{Y}}_{\gtsHVertexI}}^{\alpha},%
  {{\gtsAFOP_{Z}}_{\gtsHVertexI}}^{\alpha}\right)$
becomes an orthonormal trihedron in the continuum limit:
\begin{equation}
\label{afl/tri/XY/Z/def}
{{\gtsAFOP_{Z}}_{\gtsHVertexI}}_{\alpha}\equiv%
{\scriptstyle \frac{1}{2}}\,%
\gtsastr_{\alpha\beta\gamma}%
\left[%
{{\gtsAFOP_{X}}_{\gtsHVertexI}}^{\beta}{{\gtsAFOP_{Y}}_{\gtsHVertexI}}^{\gamma}%
-{{\gtsAFOP_{Y}}_{\gtsHVertexI}}^{\beta}{{\gtsAFOP_{X}}_{\gtsHVertexI}}^{\gamma}%
\right];
\end{equation}
inserting the expressions (\ref{afl/tri/XY})
into (\ref{afl/tri/XY/Z/def}) gives
\begin{equation}
\label{afl/tri/XY/Z/crude}
{{\gtsAFOP_{Z}}_{\gtsHVertexI}}_{\alpha}\!=\!%
{\scriptstyle \frac{2\sqrt{3}}{3\gtsSpinNorm^2}}\,%
{\scriptstyle \frac{1}{3}}\,\gtsastr_{\alpha\beta\gamma}\!%
\left[
{\gtsSpin_{\!\gtsHVertexI_{0}}}^{\beta}{\gtsSpin_{\!\gtsHVertexI_{1}}}^{\gamma}%
\!+\!{\gtsSpin_{\!\gtsHVertexI_{1}}}^{\beta}{\gtsSpin_{\!\gtsHVertexI_{2}}}^{\gamma}%
\!+\!{\gtsSpin_{\!\gtsHVertexI_{2}}}^{\beta}{\gtsSpin_{\!\gtsHVertexI_{0}}}^{\gamma}%
\right]\!.
\end{equation}
Readily,
the complex representation (\ref{afl/tri/AFDF/AFOP}) writes
\begin{equation}
\label{afl/tri/AFDF/AFOP/nested}
{\gtsAFOP_{\gtsHVertexI}}^{\alpha}=%
  {\scriptstyle\frac{1}{\sqrt{2}}}\,\gtse^{-\ci\,\gtsGP_{\gtsHVertexI}}%
  \left[{{\gtsAFOP_{X}}_{\gtsHVertexI}}^{\alpha}%
    +\ci\,{{\gtsAFOP_{Y}}_{\gtsHVertexI}}^{\alpha}%
  \right];
\end{equation}
and the pseudo-scalar representation (\ref{afl/tri/XY/Z/def}) yields
\begin{equation}
\label{afl/tri/XY/Z/nested}
{{\gtsAFOP_{Z}}_{\gtsHVertexI}}_{\alpha}=%
{\scriptstyle \frac{1}{2}}\,\ci\:%
\gtsastr_{\alpha\beta\gamma}%
\left[%
{^{\dag}\gtsAFOP_{\gtsHVertexI}}^{\beta}{\gtsAFOP_{\gtsHVertexI}}^{\gamma}
-{\gtsAFOP_{\gtsHVertexI}}^{\beta}{^{\dag}\gtsAFOP_{\gtsHVertexI}}^{\gamma}
\right].
\end{equation}
Thus, \textit{a posteriori},
for the antiferromagnetic triangular lattice,
we see that the order parameter lives on $\mathrm{SO}(3)$.
Furthermore to each hyper-particle $\left(\gtsHVertexI\right)$
we associate the chirality $\gtsAFChirality_{\gtsHVertexI}$
defined by \cite{UPTAFA}
\begin{equation}
\label{afl/tri/chirality/def}
\gtsAFChirality_{\gtsHVertexI}\equiv%
{\scriptstyle \frac{3\sqrt{3}}{2\gtsSpinNorm^3}}\:%
\gtsastr_{\alpha\beta\gamma}%
{\gtsSpin_{\!\gtsHVertexI_{0}}}^{\alpha}%
{\gtsSpin_{\!\gtsHVertexI_{1}}}^{\beta}%
{\gtsSpin_{\!\gtsHVertexI_{2}}}^{\gamma}.
\end{equation}
A straightforward calculation shows that $\gtsAFChirality_{\gtsHVertexI}$
does not depend on the local gauge field $\gtsGP_{\gtsHVertexI}$;
we get
\begin{equation}
\label{afl/tri/chirality/nested}
\gtsAFChirality_{\gtsHVertexI}=%
{\scriptstyle \frac{\gtsLStep^2}{\gtsSpinNorm}}\,%
{{\gtsAFOP_{Z}}_{\gtsHVertexI}}^{\alpha}%
{\gtsAFTR_{\gtsHVertexI}}_{\alpha}.
\end{equation}
The previous relation (\ref{afl/tri/chirality/nested})
hints strongly to introduce the nude chirality
\begin{equation}
\label{afl/tri/chirality/nude}
\gtsAFNChirality_{\gtsHVertexI}\equiv%
{{\gtsAFOP_{Z}}_{\gtsHVertexI}}^{\alpha}%
{\gtsAFTR_{\gtsHVertexI}}_{\alpha}
\end{equation}
which describes the antiferromagnetic state of the hyper-particle:
a hyper-particle ideally antiferromagnetic
(\textit{i.e.}, ``\`a la N\'eel'')
has a zero nude chirality whatever
the step $\gtsLStep$ and the quantum number $\gtsSpinDim$
may be.
The definition of the nude chirality $\gtsAFNChirality_{\gtsHVertexI}$
for any hyper-particle $\left(\gtsHVertexI\right)$
belonging to a regular crystal of any kind is obvious.
According to the relation (\ref{afl/chain/rel/chirality}),
the hyper-particles associated with
the antiferromagnetic chain have a zero nude chirality.

Let us now turn our
attention to the antiferromagnetic square lattice.
The square lattice corresponds to the plane filled
with the $2$-cube $\gtsCubicPolytope^1$ represented
in the real division algebra $\mathbb{C}$ by the quadruplet
\begin{equation}
\label{afl/square/rp/alg}
\gtsCubicPolytope^{1}=\left\{+1,+\ci,-1,-\ci\right\}.
\end{equation}
Combinations of complex conjugation and rotation $\ci$
bring the $2$-cube $\gtsCubicPolytope^1$
into coincidence with itself.
For any arbitrarily extracted hyper-crystal,
the four sites of any hyper-site $\left(\gtsHVertexI\right)$
represented respectively by
the complex numbers $+1$, $+\ci$, $-1$ and $-\ci$
are denoted respectively by
$\left(\gtsHVertexI_{0}\right)$,
$\left(\gtsHVertexI_{1}\right)$,
$\left(\gtsHVertexI_{2}\right)$ and
$\left(\gtsHVertexI_{3}\right)$.
\begin{subequations}
\label{afl/square/AFDF}
As the antiferromagnetic symmetry
is carried out by the rotation/antisymmetry~$-1$,
the local antiferromagnetic order parameter
${\gtsAFOP_{\gtsHVertexI}}^{\alpha}$ yields
\begin{align}
\label{afl/square/AFDF/AFOP}
4\,\gtsSpinNorm{\gtsAFOP_{\gtsHVertexI}}^{\alpha}&=%
\left[%
{\gtsSpin_{\!\gtsHVertexI_{0}}}^{\alpha}%
\!+\!{\gtsSpin_{\!\gtsHVertexI_{2}}}^{\alpha}%
\right]%
-\left[%
{\gtsSpin_{\!\gtsHVertexI_{1}}}^{\alpha}%
\!+\!{\gtsSpin_{\!\gtsHVertexI_{3}}}^{\alpha}%
\right],
\intertext{the complex representation
${\gtsAFAlg_{\gtsHVertexI}}^{\alpha}$ writes}
\label{afl/square/AFDF/AFAlg}
{\scriptstyle\frac{4}{\sqrt{2}}}\,%
\gtsLStep\gtsSpinNorm{\gtsAFAlg_{\gtsHVertexI}}^{\alpha}&=%
  \gtse^{-\ci\,\gtsGP_{\gtsHVertexI}}%
  \left[%
    {\gtsSpin_{\!\gtsHVertexI_{0}}}^{\alpha}%
    \!+\!\ci\,{\gtsSpin_{\!\gtsHVertexI_{1}}}^{\alpha}%
    \!-\!{\gtsSpin_{\!\gtsHVertexI_{2}}}^{\alpha}%
    \!-\!\ci\,{\gtsSpin_{\!\gtsHVertexI_{3}}}^{\alpha}%
  \right],
\intertext{and the trivial representation
${\gtsAFTR_{\gtsHVertexI}}^{\alpha}$ reads}
\label{afl/square/AFDF/AFTR}
4\,\gtsLStep^2{\gtsAFTR_{\gtsHVertexI}}^{\alpha}&=%
  {\gtsSpin_{\!\gtsHVertexI_{0}}}^{\alpha}
  \!+\!{\gtsSpin_{\!\gtsHVertexI_{1}}}^{\alpha}
  \!+\!{\gtsSpin_{\!\gtsHVertexI_{2}}}^{\alpha}
  \!+\!{\gtsSpin_{\!\gtsHVertexI_{3}}}^{\alpha}.
\end{align}
\end{subequations}
\begin{subequations}
\label{afl/square/rel}
In the classical limit the constraints obey:
\begin{gather}
\label{afl/square/rel/unit}
{\gtsAFOP_{\gtsHVertexI}}^2=%
1-{\scriptstyle\gtsLStep^2}\;%
{\gtsAFAlg_{\gtsHVertexI}}^{\alpha}{^{\dag}\gtsAFAlg_{\gtsHVertexI}}_{\alpha}
-{\scriptstyle\frac{\gtsLStep^4}{\gtsSpinNorm^2}}\;%
\gtsAFTR_{\gtsHVertexI}^2,\\
\label{afl/square/rel/ortho}
{\gtsAFOP_{\gtsHVertexI}}^{\alpha}{\gtsAFAlg_{\gtsHVertexI}}_{\alpha}=%
-{\scriptstyle\frac{\gtsLStep^2}{\gtsSpinNorm}}\,%
{\gtsAFAlg_{\gtsHVertexI}}^{\alpha}{\gtsAFTR_{\gtsHVertexI}}_{\alpha},\\
\label{afl/square/rel/chirality}
{\gtsAFOP_{\gtsHVertexI}}^{\alpha}{\gtsAFTR_{\gtsHVertexI}}_{\alpha}=%
{\scriptstyle\frac{\gtsSpinNorm}{4}}\,%
\left[%
{\gtsAFAlg_{\gtsHVertexI}}^{\alpha}%
  {\gtsAFAlg_{\gtsHVertexI}}_{\alpha}%
+{^{\dag}\gtsAFAlg_{\gtsHVertexI}}^{\alpha}%
  {^{\dag}\gtsAFAlg_{\gtsHVertexI}}_{\alpha}%
\right].
\end{gather}
\end{subequations}
Hence
the number of degrees of freedom (equals~to~$8$) is unchanged
and the order parameter ${\gtsAFOP_{\gtsHVertexI}}^{\alpha}$
is effectively normalized to unity in the continuum limit.
The nude chirality $\gtsAFNChirality_{\gtsHVertexI}$
for any hyper-particle $\left(\gtsHVertexI\right)$
belonging to the antiferromagnetic square lattice
is obtained from (\ref{afl/square/rel/chirality}):
\begin{equation}
\label{afl/square/chirality/nude}
\gtsAFNChirality_{\gtsHVertexI}=%
{\scriptstyle\frac{\gtsSpinNorm}{4}}\,%
\left[%
{\gtsAFAlg_{\gtsHVertexI}}^{\alpha}%
  {\gtsAFAlg_{\gtsHVertexI}}_{\alpha}%
+{^{\dag}\gtsAFAlg_{\gtsHVertexI}}^{\alpha}%
  {^{\dag}\gtsAFAlg_{\gtsHVertexI}}_{\alpha}%
\right].
\end{equation}
Henceforth,
if each hyper-particle is ideally antiferromagnetic
($\gtsAFNChirality_{\gtsHVertexI}\simeq 0$),
in the continuum limit
${\gtsAFAlg_{\gtsHVertexI}}^{\alpha}$
becomes an orthogonal dihedron and the pair
$\left({\gtsAFAlg_{\gtsHVertexI}}^{\alpha},%
  {\gtsAFOP_{\gtsHVertexI}}^{\alpha}\right)$
an orthogonal trihedron:
the order parameter lives then in $\mathrm{SO}(3)$,
as for the triangular lattice.

\textit{A fortiori},
for both the antiferromagnetic triangular and square lattices
in the continuum limit,
the local order parameter appears to be
an orthogonal trihedron defined up to
a rotation around the pseudo-scalar axis:
the ``antiferromagnetic'' field interacts with a local gauge field.
For such interacting fields the Lagrangian formulation
leads to the nonlinear $\sigma$-model (as anticipated)
with an electromagnetic-field-like term
and a Hopf term \cite{HSTL}.
The coupling strengths are expected to emerge during the passage
to a continuum field theory
from a discrete field theory governed
by an effective Hamiltonian like (\ref{Hm/Heiseinberg/afnn})
using our particle-like formalism \cite{HSTL}.

In conclusion, we have outlined
the construction of novel local antiferromagnetic degrees of freedom
for low dimensional antiferromagnetic lattices
($\gtsDimension=1,2,3,4$).
The Haldane decomposition for the antiferromagnetic chain
has been recovered.
Both the antiferromagnetic triangular and square lattices
have been shown to exhibit in the continuum limit
a similar local order parameter
living on $\mathrm{SO}(3)$ and
interacting with a local gauge field:
the corresponding Lagrangian formulation
presents a nonlinear $\sigma$ term and a Hopf term.
Besides, the construction reproduces
the real division algebra hierarchy
fulfilled by the nonlinear $\sigma$-model.
To recap, we have outlined
a novel extension of the spin group by a local gauge field.


\end{multicols}
\end{document}